# Once Upon an AI: Six Scaffolds for Child-AI Interaction Design, Inspired by Disney

**Dr. Nomisha Kurian**
**University of Warwick**
nomisha-chandran.kurian@warwick.ac.uk

## Abstract

To build AI that children can intuitively understand and benefit from, designers need a design grammar that truly serves their developmental needs. This paper bridges Artificial Intelligence design for children - an emerging field still defining its best practices - and children's animation, a well-established field with decades of experience in engaging children through accessible and engaging storytelling. Pairing Piagetian developmental theory with design pattern extraction from 52 works of children's animation, the paper presents a "six scaffold" framework that integrates design insights transferable to child-centred AI design: **(1) *signals* for visual animacy and clarity, (2) *sound* for musical and auditory scaffolding, (3) *synchrony* in audiovisual cues 4) *sidekick*-style personas (5) *storyplay* that supports symbolic play and imaginative exploration,** and **(6) *structure* in the form of predictable narratives**. These strategies - long refined in children's animation - function as multimodal scaffolds for attention, understanding, and emotional attunement, supporting children's learning and emotional comfort. This structured design grammar is transferable to AI design. By reframing cinematic storytelling and child development theory as design logic for AI, the paper offers heuristics for crafting intuitive AI that aligns with children's cognitive stages and emotional needs. The work contributes to design theory by showing how sensory, affective and narrative techniques can inform developmentally-attuned AI design. Future directions include empirical testing, cultural adaptation, and participatory co-design.

## 1. Introduction

As AI enters children's lives - from educational apps and smart speakers to generative tools - research stresses the need for alignment with children's cognitive, socio-emotional, and physical needs (Chen, 2025; Kurian, 2024; Wang et al., 2022). This paper argues that "developmentally aligned AI" (Kurian, 2025) can draw not only on cognitive science and usability research but also on the genre of animation, which has spent decades crafting cognitively accessible, emotionally engaging experiences for children (Ratelle, 2018). Using developmental psychology, it translates cinematic strategies into AI design heuristics for 2-11 year olds.

**An important ethical caveat**: this paper does not endorse companion-style AI for children that fosters emotional dependency, unbounded intimacy, or other ethically problematic attachments. It focuses on functional AI for tightly boundaried tasks such as language learning, and the interaction design of systems (e.g. educational tutoring agents) that children engage with directly through speech, gestures, or shared activities. This excludes non-interactive AI such as algorithmic recommendation systems. Interaction design encompasses both communication and visual presentation - conversational style, tone, behavioural cues, and visual form (facial features, body language, and aesthetic) - as these shape how children perceive, engage with, and learn from AI, and underpin the six scaffolds proposed. Children's animation offers an underused source of design knowledge here. For decades, animated media has been crafted to match young viewers' perceptual strengths, attention patterns, and emotional needs, delivering content that is both engaging and developmentally appropriate (Ratelle, 2018; Wells, 2002).

Although AI and animation operate in different domains, both must capture attention, convey messages clearly, and scaffold understanding through visual, auditory, and narrative cues. They differ in two key respects. Animation is fixed: all viewers receive the same scripted sequence, allowing precise control over pacing, tone, and framing. AI must operate in real time, interpreting ambiguous child input, making rapid decisions, and adapting to changing cognitive and emotional states. Animation can perfect synchrony through editing; AI must achieve it dynamically.

These differences create both challenges and opportunities. Animation's "rich grammar" of visual, auditory, and narrative techniques - honed to support early cognition - can inform AI's interactive demands, while AI's responsiveness and personalisation can extend animation's strengths by tailoring cues to each child's moment-to-moment state.

The need for such cross-pollination is pressing as child-facing AI expands into voice assistants, embodied robots, and conversational tools. Current systems often suffer from sensory overload, inconsistent affective feedback, weak narrative coherence, and poor verbal–visual synchrony (Druga et al., 2017; Garg & Sengupta, 2020; Kurian, 2024). Achieving "developmentally aligned AI" (Kurian, 2025) requires more than technical optimisation: it demands scaffolding for affective clarity, predictable structure, and multimodal integration, mirroring the strengths of children's media while meeting AI's real-time demands.

Accordingly, this paper draws on developmental psychology to derive "six scaffolds" from animation: design heuristics for supportive and adaptive child-AI interaction:

1. Signals: Treating visual cues as cognitive supports for children still learning to infer emotional states. This includes bold, readable gestures and expressions - slow eye movements, high-contrast posture changes, and prolonged gaze - giving children time to perceive and interpret social signals.

2. Sound: Using sound and music as core interactive modalities that scaffold comprehension and narrative flow. Audio motifs can anchor attention, convey affective shifts, and shape turn-taking or transitions – particularly valuable for children developing executive function and benefiting from rhythmic, multimodal cues.

3. Synchrony: Building reassuring interactional environments through cross-modal consistency – soothing movements paired with gentle musical phrases, or visual affirmations aligned with warm vocal feedback – to offer perceptual reassurance and emotional legibility.

4. Sidekicks: Adopting a collaborative, non-hierarchical persona – a playful, encouraging "sidekick" showing humility and trial-and-error, rather than a directive instructor. This emphasises dialogue and shared exploration, and helps children develop more accurate mental models of AI.

5. Storyplay: Using AI's generative capacity to support open-ended, co-constructed play: allowing the child to improvise and the system to riff back, turning sessions into shared, imaginative "storyplay" rather than scripted tasks. This aligns with theories of symbolic play as a developmental zone for higher-order reasoning, role-taking, and problem-solving.

6. Structure: Applying stable patterns and predictable interaction structures to reduce cognitive load while adjusting challenge levels to stretch skills. This rhythm of familiarity and growth supports progression by ensuring the child knows what to expect while being gently pushed beyond current capacities.

This paper makes three contributions: framing cinematic storytelling as a design vocabulary for AI; synthesising media studies, developmental psychology, and child–computer interaction; and offering transferable heuristics for creating intuitively understandable AI for children.

# Background

### 2.1 Child-Computer Interaction

Over two decades, CCI has developed heuristics on usability, safety, and cognitive scaffolding (Druin, 2002; Chiasson & Gutwin, 2005; Nilsson et al., 2020; Markopoulos et al., 2008). Guidelines stress intuitive navigation, low cognitive load, and alignment with children's perceptual and motor abilities (Hourcade, 2008). For young children, visual simplicity, minimal text, and familiar characters are advised (Markopoulos et al., 2008), while research into tangibles, voice input, and playful modalities supports emerging symbolic and social cognition (Antle, 2007).

Most prior work centres on GUI design or physical computing, but new strategies are needed for socially interactive AI, including smart speakers and voice assistants (Andries & Robertson, 2023) and generative or conversational tools for children (Druga et al., 2018; Garg & Sengupta, 2020). AI for children ranges from tangible supports - peer-tutor robots and responsive toy agents (de Haas et al., 2022; Hsiao et al., 2015; Kewalramani et al., 2021) - to intangible systems such as voice assistants for social interaction and information retrieval (Aeschlimann et al., 2020; Girouard-Hallam & Danovitch, 2022; Xu et al., 2022). It also includes adaptive educational tools in interactive music and game environments (Addessi & Pachet, 2005; de Castro Rodrigues et al., 2022; Bonneton-Botté et al., 2020) and AI as a learning subject, where children teach or explore AI concepts (Gulz et al., 2020; Su & Yang, 2023). Across these applications, AI offers real-time responsiveness, adaptive dialogue, and affect simulation through voice, gesture, and pacing - capabilities that can lead children to perceive AI as a social entity (Chubb et al., 2022; Kory-Westlund & Breazeal, 2019; Turkle, 2011), making developmental alignment crucial (Kurian, 2025).

### 2.2 Affective Computing

Affective design in pedagogical agents has been shown to increase motivation and learning (Lester et al., 1997; Moreno et al., 2001). The "persona effect" suggests that agents perceived as expressive and friendly improve engagement, even when instructional content is identical (Lester et al., 1997). Studies show children prefer agents with facial expressiveness and vocal prosody (Chiasson & Gutwin, 2005), and that consistent social cues such as smiling and vocal praise support learning outcomes (Kory-Westlund et al., 2018).

Still, despite the growing presence of AI systems in children's lives, current design practices remain poorly attuned to developmental needs. Recurring challenges have been noted by scholars. First, insights from child psychology remain under-utilised in AI design (Rudenko et al. 2024) and emotional legibility remains low, with minimal or mismatched affective cues that make it difficult for children to interpret AI intentions (Druga et al., 2017; Kory-Westlund & Breazeal, 2019). A "significant research gap" exists around tailoring AI "facial" design to children's preferences, for example (Oliva et al. 2025, p. 1). Second, AI interactions often lack predictable, scaffolded structures, generating content dynamically but without the stable rhythms that help children orient themselves (Xu & Warschauer, 2019). Third, there is little narrative coherence, with many systems offering pre-scripted, static and transactional exchanges rather than story-driven contexts that aid comprehension and engagement (Chubb et al., 2022). Fourth, trust calibration is problematic: anthropomorphic cues can lead to over-reliance, while unexplained errors can erode trust (Mehrotra et al., 2024; Kurian, 2024). Fifth, AI outputs often lack cultural and developmental appropriateness, producing language, pacing, or scenarios that fail to resonate - or even pose risks (Wang et al., 2024). Finally, multimodal synchrony is weak, with a need for audio, visual, and interactive cues to be presented in smooth and cognitively manageable ways (Antle, 2007; Aslan et al., 2024). These gaps point to the absence of a developmentally grounded, multimodal design grammar in child-AI design. The six-scaffold framework proposed here draws on children's animation to supply such a grammar - offering clear, narratively coherent, and culturally adaptable interaction patterns that align with children's ways of sensing, storytelling, and meaning-making.

### 2.3 Children's Media

Children's animation has long prioritized age-appropriate aesthetic and narrative techniques that align with children's cognitive and perceptual abilities (Calvert & Wilson, 2009) and research exploring what younger children like or dislike in character styles (Carter & Eckler, 2016). Decades of production refinement have honed audiovisual conventions to support children's comprehension; for instance, segmenting action with brief fades or sound cues to help 3–6-year-olds detect narrative boundaries (Anderson & Pempek, 2000); syncing music and voice-over rhythms to bolster recall and engagement (Mares & Pan, 2013); and using consistent audio motifs - such as signature jingles before and after learning segments - to assist preschoolers in internalizing program structure and improve content retention (Fisch, 2004; Linebarger & Piotrowski, 2009). Animation thus hosts well-established design techniques to reduce children's cognitive load, regulate mood, and scaffold emerging learning processes.

Disney films are particularly known for consistent tonal signalling and use of bold visual and auditory cues (Wells, 2002; Tremosa, 2024). These map directly onto developmental capacities: preschoolers can track emotional arcs and motives when cues are overt (Wilson & Drogos, 2007), and can understand and recall better when exaggerated facial expressions and musical motifs reinforce meaning (Calvert & Kotler, 2003). Developmentally-aligned design offers what might be called *intuitive legibility*: children infer emotional or narrative shifts with little need for instruction.

This intuitive legibility can enhance AI design. Kurian and Saad (2024) describe how digital storytelling offers "arrows of imagination" (p. 151) that pull children into co-constructed narrative spaces through gesture, voice, and play. Intuitively understandable AI interactions can encourage a child to play and explore, shifting their experience from passive consumption to participatory storytelling.

### 2.4 Early Childhood Development

In early childhood, rapid brain maturation occurs in sensorimotor, attentional, and emotional-regulation circuits (Su & Yang, 2024). Piaget's stages of cognitive development (1952) shift from sensorimotor (0–2 years) to preoperational (2–7 years), concrete operational (7–11 years), and formal operational (11+ years). In the preoperational stage, children show symbolic thinking, limited logical reasoning, and reliance on perceptual cues. They engage in imaginative play, personify objects, and favour concrete over abstract representations, responding to visual salience, rhythmically structured sound, expressive characters, and predictable narrative scaffolds (Flavell et al., 1995). Repetition, musical motifs, and verbal signposts further scaffolds comprehension and predictability (Addessi & Pachet, 2006; Anderson et al., 2000). Without mature metacognitive and digital-literacy skills, preschoolers may anthropomorphise AI and overtrust its outputs, risking confusion or misplaced reliance (Kurian, 2023; Su, Ng, & Chu, 2023; Sun et al., 2024). Misaligned AI - opaque feedback, poor recognition of children's speech, unexpected interruptions, or overstimulation - can disrupt self-regulation and attachment, fuelling anxiety and disengagement (Kurian, 2024). By contrast, "developmentally aligned" AI (Kurian, 2025) leveraging tangible interfaces, adult-mediated scaffolding, and play-based narratives can support cognitive growth and socioemotional security (Delaney & Chen, 2025), offering predictable, meaningful interactions that foster curiosity and agency (Su & Yang, 2024; Su et al., 2023).

In contrast, children in the concrete operational stage begin to master logic, understand how to classify objects, and show improved perspective-taking - yet their reasoning remains grounded in tangible, structured experiences (Inhelder & Piaget, 1958). For this group, AI can introduce more complex cause-and-effect relationships, gently increase abstraction, and scaffold reflection through conversational prompts or feedback loops. Across both stages, children benefit from systems that balance novelty with structure, and that communicate meaning through multimodal, perceptually accessible cues - making developmental sensitivity crucial to designing AI that supports both play and learning.

### 2.5 Gaps in Cross-Domain Translation

Despite overlaps across HCI, CCI, and media studies, few works systematically translate animated media strategies into child–AI interaction. While media scholars have examined children's responses to music, story, and visuals (Napoli, 2001; Wright et al., 2001), these insights rarely guide AI design. Lacking a structured bridge, teams may imitate surface traits from popular culture ("make it friendly like Baymax") without applying the narrative grammar and cross-modal logic that make such characters engaging and legible. Early AI work drew on media inspiration - Breazeal (2004) referenced fictional robots' "rich personalities," and Cavazza et al. (2001) and Kory-Westlund & Breazeal (2019) used gestural or animation-style movement - but few studies provide detailed guidance from children's animation. This paper addresses that gap by mapping visual, auditory, affective, and narrative techniques from animated films into structured design heuristics for child–AI interaction.

## 2. Methods

This paper is part of a larger mixed-methods project on child-centred AI design (2023–2025), premised on the idea that AI for children need not start from scratch but can adapt mature design grammars from media such as television, film, picture books, and video games, refined over decades to meet children's cognitive, emotional, and social needs. These media offer established repertoires of visual, auditory, and narrative strategies suited to interactive contexts. The project had three phases:

(1) mixed-methods analysis of media content

(2) semi-structured interviews with CCI professionals and designers

(3) participatory co-design sessions with educators, children, and early childhood experts.

This paper reports on phase one, which extracted developmentally appropriate strategies from ~70 hours of footage across 52 Walt Disney Animation Studios films (1937–2021), chosen for their historical breadth and culturally influential storytelling grammar. A qualitative thematic analysis identified recurring sensory, narrative, and affective techniques, which were then mapped to developmental and CCI design frameworks. Building on earlier thematic categories - emotional clarity, character relatability, and narrative rhythm - this strand systematised findings by translating cinematic techniques into design-relevant scaffolds through Piagetian theory and child-AI literature. The resulting six scaffolds refine the broader project's findings, offering structured, transferrable heuristics for designing intuitive, legible AI for children aged ~2–11.

### 3.1 Data Selection

This analysis examined 52 films from Walt Disney Animation Studios (WDAS), chosen for their commercial and cultural impact in children's media (Tremosa, 2024). 37 were from the 1937–1999 "Classic" era - often called Disney's Golden Age[1] and Renaissance due to artistic and technical innovation - while the remaining 15, like *Frozen* (2013), were included for their contemporary influence (Wells, 2002). Together, the films offered a longitudinal view of animation crafted to capture attention and engage viewers across early and middle childhood.

### 3.2 Thematic Analysis

This study used a thematic analysis (Braun & Clarke, 2006) of the 52 films to identify media strategies that could inspire developmentally appropriate AI interaction design. From the outset, Piagetian developmental theory and CCI literature were used in tandem to guide the analytic process. Piaget's cognitive stage theory offered a developmental lens, helping to identify child-centred perceptual, cognitive, and emotional needs, while CCI research contributed practical design principles for digital interactions for children.

The initial codebook was developed through an inductive-deductive approach, drawing on a subset of 15 animated films[2] and six key works[3] from the CCI literature. The films were chosen to span classic and contemporary aesthetics, cross-cultural settings, and character types, offering a broad view of media techniques across eras.

Coding of this pilot set in NVivo included scene-level annotation of elements judged to be potentially relevant for child-AI design. The resulting codebook included: **Expressiveness** – exaggerated voice, facial, or body cues; **Narrative Structure** – clear arcs and predictable sequencing; **Audio/Music Cues** – sound to mark transitions, signal emotion, or hold attention; **Environment & Visual Style** – visual/spatial cues for comfort, threat, or transition; **Sidekick Dynamics** – support, humour, or guidance from secondary characters; **Play & Agency** – sequences inviting imaginative exploration or interactive choice.

Analysis of all 52 films refined the codebook through iterative comparison, merging overlaps (e.g., emotional vs. rhythmic musical cues) and adding categories such as guidance/scaffolding from scenes where characters coached protagonists. Coding in NVivo used the final codebook and analytic memos linking media strategies to developmental affordances and AI design implications - for instance, a sidekick modelling coping was tagged as both emotional scaffolding and a cue for affectively responsive AI. Axial coding then grouped related codes into cross-cutting developmental patterns, each interpreted through Piagetian theory and CCI literature.

---

[1] The "Classic"-era films included *Snow White and the Seven Dwarfs* (1937), *Cinderella* (1950), *The Little Mermaid* (1989), *Beauty and the Beast* (1991), *Aladdin* (1992), *Pocahontas* (1995), and *Mulan* (1998). This 1937-1999 "Classic" era of WDAS encompasses four recognised sub-eras: the *Golden Age* (1937-1942), *Silver Age* (1950-1967), *Bronze* or *"Dark" Age* (1970-1988), and the *Disney Renaissance* (1989-1999). The *Package Era* films (1943-1949) are also included, as they fall within this chronological span and reflect the studio's wartime and post-war production responses. These sub-eras represent distinct phases in Disney's stylistic evolution, characterised by shifts in animation technique, sound design, narrative tone, and technological experimentation

[2] *Snow White* (1937), *Pinocchio* (1940), *The Lion King* (1994), *Toy Story* (1995), *Mulan* (1998), *Finding Nemo* (2003), *The Incredibles* (2004), *Up* (2009), *Tangled* (2010), *Frozen* (2013), *Big Hero 6* (2014), *Zootopia* (2016), *Moana* (2016), *Coco* (2017), and *Encanto* (2021)

[3] This included: Hanna, Risden, and Alexander (1999) on design principles for different developmental stages; Chiasson and Gutwin (2005) on patterns of interaction and interface conventions for children; Hourcade (2008) on child development in human-computer interaction; Druin (2002) on cooperative inquiry and participatory design with children; and Antle (2007) on tangible interfaces and children's conceptual understanding. These seminal works emphasise child-facing design strategies—multimodal cues, visual clarity, playful interaction, scaffolding, and participatory structure—used to shape initial coding categories.

- **Visual expressiveness and clarity** was linked to preoperational children's limited theory of mind and need for concrete relational cues (Piaget, 1951).

- **Musical and auditory scaffolding** was interpreted as a form of rhythmic, multimodal support for attention and emotion regulation, aligning with limited working memory and sensory integration needs (Inhelder & Piaget, 1958).

- **Audiovisual synchrony for emotional comfort** aligned with preoperational and early concrete operational stages, supporting children's emotional sense-making through layered sensory cues.

- **Sidekick-style companion guidance** aligned with Piaget's (1951) constructivism: provoking thinking without over-directing, and using brief, stage-appropriate prompts to help children reorganise their understanding while maintaining autonomy.

- **Symbolic play and imaginative exploration** reflected the hallmark behaviours of the preoperational stage, such as pretend scenarios, perspective-taking, and personification (Wellman, 2014).

- **Predictable and scaffolded interaction structure** supports the needs of children entering the concrete operational stage, helping them manage sequential logic and learn through guided routines.

This developmentally-informed coding process helped ensure that interpretations were both descriptive of children's media and actionable for AI design.

### 3.4 Study Limitations

To strengthen credibility, preliminary themes were reviewed by an interdisciplinary group including child development specialists; for instance, "Audiovisual Synchrony" was reframed to emphasise emotional safety. A CCI colleague also conducted peer debriefing through coding sample review. Bias was mitigated by focusing on documented animation techniques (Johnston & Thomas, 1981) and grounding interpretations in developmental theory (Piaget, 1952). The study remains interpretive, however, and shaped by a developmental lens. Alternative readings are possible. A key limitation is the absence of direct child involvement at this stage; these heuristics require empirical validation, but are rooted in Disney animation's long-established capacity to engage children across developmental stages.

While the proposed scaffolds apply to both screen-based avatars and physically embodied robots, the two modalities diverge in ways that affect implementation. Embodied robots face hardware constraints - limited degrees of freedom, motor noise, battery life - and contextual factors such as physical safety and shared space negotiation. A screen-based avatar can render a wide range of facial expressions via 2D or 3D animation, whereas a robot must approximate them mechanically or through LED displays, potentially reducing nuance or introducing latency. Likewise, tablets can synchronise voice and gesture fluidly, but robots must coordinate speech with motor actuation, risking desynchronisation that disrupts coherence. Design emphasis should therefore adapt: robots may require slower, exaggerated gestures, careful positioning, and tactile cues, while screen-based agents can use rapid visual transitions, detailed facial expressivity, and layered audiovisual effects. Future research should address these modality-specific differences, particularly in mixed-reality settings where both agent types coexist.

# 3. Findings and Discussion

## 4.1 Signals: Visual Expressiveness and Clarity

**Disney and Child Development**

Disney's animation style, built on principles like *exaggeration*, *appeal*, and *staging* (Tremosa, 2024), provides a blueprint for visual legibility. Emotions are painted in broad strokes: Cinderella's sadness appears as large tears; Ariel's wonder gleams through widened eyes and bouncing gestures; Jafar's fury crackles in every contorted line of his face. Side characters show exaggerated emotions, too - Olaf flails in panic, Mushu shrinks in embarrassment. Crucially, the camera-work pairs these expressions with deliberate *visual pacing*: facial reactions linger, gestures resolve slowly, and the camera often holds a beat before or after speech. This narrative breath gives children the time they need to process emotion, intention, and social cues.

These techniques signal emotions in ways matched to the perceptual strengths of Piaget's preoperational stage (ages 2–7). With emerging but incomplete theory of mind, children rely on clear cues - facial expression, vocal tone, gesture - rather than inferring causes, and struggle with subtle or ambiguous affect (Bayet et al., 2018; Wellman, 2014). Weak abstract reasoning means emotional legibility must come through clarity and pattern. Disney's expressive animation offers a scaffold, modelling how to externalise internal states in forms children can reliably decode.

**Design Implications**

AI characters - whether screen-based avatars or embodied robots - can move beyond surface-level friendliness by mirroring this cinematic grammar. While commercial robots like

Moxie pair face recognition with pre-scripted animations, systems selecting expressions dynamically from child-affect models are rare, and empirical work on adaptive facial-gesture expressivity in child–AI interaction is limited. Some results are promising: in a within-subjects study, primary school children enjoyed interacting more with an emotionally expressive robot than with a non-expressive one (Tielman et al., 2014). Similarly, in a two-week vocabulary-tutoring study, children retained more target words and engaged more when a tablet-faced robot gave contingent, expressive feedback after correct responses (Ahmad et al., 2019). These findings suggest that even minimal expressive adaptability in simple rule-based systems can meaningfully shape interaction.

Building on this, AI can translate Disney-style expressiveness into contingent social feedback by sensing a child's actions and giving a clearly legible visual response. Developmentally attuned expressiveness may use bold, readable gestures and expressions - slow eye movements, high-contrast posture changes, prolonged gaze, and overt gestures with pauses to avoid cognitive overload, especially for young or neurodivergent users. Large eyes, exaggerated gestures, and synchronised voice tone can help preoperational children decode intent. In practice, this can involve an affect-state controller (e.g., finite-state machine or behaviour tree) mapping observations - Automatic Speech Recognition confidence, pauses, gaze-off-screen, repeated requests - to display behaviours such as a "thinking" face, head tilt, or self-talk, with dwell time for children to read the signal. Timing responses in real time to match the child's behaviour is important: MIT's Tega times facial animations and back-channel gestures to a child's speech pauses and turn-taking, and in one study children spoke more energetically, made more eye contact, and perceived the contingent robot as more attentive than a non-contingent one (Park et al., 2017). Similarly, robots adapting speech delivery to a child's rhythm have elicited more positive emotion and stronger rapport in storytelling (Kory-Westlund & Breazeal, 2019).

Implementation faces challenges: real-time rendering and affective speech synthesis may not produce natural yet clear expressions across devices or bandwidths, and responses to expressiveness vary - some neurodiverse users may find cues overwhelming, while others rely on them. A practical approach could be adjustable intensity (low, medium, high) set by caregivers or in-session adaptation, without storing profiles, to protect privacy while tailoring interaction.

Beyond moment-to-moment contingency, AI can stabilise and personalise mappings between internal states and visible cues across repeated interactions. In Disney films, exaggerated expressions are reinforced by narration or music; similarly, AI can use predictable cues - "sparkle eyes + smile" for success, "slumped posture + gentle blink" for misunderstanding - while adjusting intensity and pacing to each child. A robot that slumps when it fails and perks up when successful creates stable, interpretable patterns children can internalise. Unlike static systems, AI can tailor cues, learning that one child prefers visual praise and another verbal encouragement. Success moments might pair sparkling eyes, animated claps, or a beaming face with synchronised congratulatory audio, tuned in real time to the child's actions. This adaptive expressiveness builds a shared "affective grammar".

## 4.2 Sound: Musical and Auditory Scaffolding

**Disney and Child Development**

Since young children make sense of the world through perceptual cues rather than abstract reasoning, sound can be a powerful temporal and emotional guide (Addesi & Pachet, 2006). When reading is still emerging, children rely on rhythm, tone, pitch, and musical motifs to track what is happening, how to feel about it, and what comes next (Register & Humpal, 2007).

Disney films leverage sound to great effect. Leitmotifs, key changes, and abrupt silences cue narrative transitions with precision children can perceive. In *Snow White*, cheerful singing in the dwarfs' cottage gives way to tense, minor-key music when the Queen arrives disguised as a hag - immediately cueing danger. A sudden drop to silence or a shift in tone when a villain appears is a recurring device that even very young viewers understand.

Disney also uses music to mark transitions in time, space, or tone. In *The Lion King*, Simba's growth is compressed into a musical montage set to "Hakuna Matata." In *Cinderella*, a whimsical chime accompanies the Fairy Godmother's appearance - pulling even distracted viewers back in. In *The Little Mermaid*, "Under the Sea" transitions us into a fantastical carnival scene. In *Frozen*, Elsa's motif recurs when her emotions surge, helping children infer a connection between music, mood, and action. Smaller auditory cues are equally meaningful: in *Peter Pan*, the shimmering chime of pixie dust becomes a sonic shorthand for flying, while the ticking clock reliably signals the crocodile's approach without needing visual confirmation. Even these brief, non-musical cues act as temporal anchors, alerting children to character proximity or plot progression.

**Design Implications**

Disney's mastery of sound design - using music and audio motifs to guide attention, mark transitions, and cue emotion – can inform child-AI design. In animation, leitmotifs, tonal shifts, and silence scaffold perception for children still developing abstract reasoning (Ratelle, 2018; Wells, 2002). AI can adopt a similar auditory grammar with a key advantage: real-time responsiveness. Unlike fixed scores, it can generate, modulate, and personalise sound dynamically in response to a child's input or emotional state (Dash & Agres, 2024). For instance, if a child hesitates, the system might soften prosody, lower background volume, or play a calming motif.

Though auditory aspects of human-machine interfaces are well-studied (see Nees & Liebman, 2023) and music-based interventions do exist for children with autism (Feng et al., 2022), AI-driven adaptive sound for children remains rare. One example is Faria and Ayrosa's (2025) use of Bayesian Immediate Feedback Learning (BIFL), a real-time framework that adaptively selected auditory, visual, or textual cues based on children's EEG signals and facial expressions. In a four-week trial with 40 neurodivergent children, their system improved concentration by 22.4%, emotional engagement by 24.8%, and game performance by 32.1%, with large effect sizes (Cohen's d > 1.4). Notably, auditory cues were especially effective for children with ADHD (p. 24), suggesting how dynamic, context-aware sound can act as a developmental scaffold in AI-mediated interaction.

Implementation can range from rule-based loops (e.g., high-pitched jingle for correct answers; minor-key "uh-oh" after repeated errors) to real-time modulation adjusting pitch, tempo, or volume based on voice or behaviour. Rising fatigue might trigger brightened background audio; long silences could prompt a soft "Ready when you are!" tune. Alexa Reading Sidekick slows TTS prosody to sustain attention (Sokolova, 2021), and Kory-Westlund and Breazeal (2019) found that robots matching children's pitch and tempo increased joy, vocabulary use, and compliance.

Such sonic scaffolding is vital for pre-readers who process auditory cues before mastering text (Choudhury & Benasich, 2011; Hanna et al., 1999). Synchronised sound and visuals aid memory, focus, and comprehension, reducing cognitive load (Piaget, 1952; Nilsson et al., 2020). Disney often layers channels, as in *The Little Mermaid*, where Ariel and her friends name her undersea artefacts aloud ("This is a dinglehopper!"), hold them up to the camera, and underscore the moment with music - ensuring that even if a child's attention drifts, at least one sensory channel will carry the narrative cue. Voice-based AI can similarly signal mode changes through coordinated voice and sound, improving understanding (Liu, 2018) and engagement (Chaspari & Lehman, 2016).

However, cues must be emotionally congruent and cognitively manageable. Overuse or mismatched tones can overwhelm children, while rhythmic, bounded sounds punctuating success, transitions, or assistance are more effective (Addessi & Pachet, 2006; Register & Humpal, 2007). Disney uses music sparingly to highlight key beats, a strategy AI can emulate to establish predictable "beginning-middle-end" patterns.

Integrating responsive, context-aware audio is technically challenging, as existing systems often rely on static soundbanks rather than adaptive composition. Privacy-sensitive adaptation could occur by detecting in-session behaviours (e.g., prolonged hesitation triggering a reassuring sound) without logging this data beyond the session. Designers should also be aware of cultural differences in music perception and ensure cues are appropriate for diverse audiences - participatory testing with children from varied cultural backgrounds can help fine-tune these auditory scaffolds. Finally, while AI can personalise sound cues over time, such personalisation should remain adjustable, allowing caregivers or children to select preferred sound intensity or frequency to accommodate neurodiverse needs and prevent sensory overload. In sum, audio cues in child–AI can serve as developmental scaffolds - responsive, personalised, and co-regulatory - shaping interactions that are natural, supportive, and rich.

## 4.3 Synchrony: Audiovisual Consistency

**Disney and Child Development**

For younger children in unfamiliar or demanding situations, emotional clarity and sensory coherence are essential. They depend on concrete, perceptual input to interpret emotions and regulate feelings (Piaget, 1951; Inhelder & Piaget, 1958), yet their ability to reconcile conflicting cues is limited (Brady et al., 2024). Digital systems that align warm colours, soft animations, gentle voices, and familiar patterns can enhance comfort, predictability, and emotional safety.

Disney films model this principle of multimodal coherence with care. Scenes coded as “safe” - such as Snow White’s pastel cottage - are rarely safe through visuals alone - they are underscored by soothing background music, playful sound effects, soft transitions, and voice acting that mirrors calm. These design elements act in tandem. For instance, when Belle is attacked by wolves in Beauty and the Beast, the peril is short-lived and quickly followed by a transition to a warmly lit, firelit room accompanied by gentle music. This kind of resolution helps children experience intensity without becoming overwhelmed. It models a regulated emotional arc: tension rises, support appears, and calm is restored - developmentally critical support for child-viewers still learning to manage their own emotions.

**Design Implications**

AI can surpass animation by delivering adaptive synchrony in real time, adjusting visuals and audio to a child’s mood, pace, or performance. Beyond synchronising cues, it can modulate them moment-to-moment - for instance, if disengagement is detected through reduced speech or gaze, the system might soften animations, dim lighting, and slow music to re-regulate affect. Using multimodal sensing (camera, mic, input timing), AI can maintain cross-modal coherence, offering a digital analogue to co-regulation by adjusting tone, pacing, and expression to sustain a safe, legible environment. Coherence is crucial: mismatched cues (e.g., friendly face with robotic voice) can confuse children still learning to integrate sensory input. Repeating consistent sound–image pairings - success chime + smile, jingle + wave for transitions - helps children form intuitive associations. Over time, this synchrony becomes an affective grammar: a stable cue set signalling feedback, transitions, or reassurance.

Implementing adaptive synchrony poses technical and inclusivity challenges. Affect sensing via camera, microphone, or other sensors can be error-prone in home settings with variable lighting, noise, or hardware, and cross-device performance may vary - high-end devices render smooth, synchronised output, while lower-spec ones risk lag or desynchrony. Inclusivity is also key: some children, particularly with sensory sensitivities, may find full audiovisual synchrony overstimulating. Designers can mitigate this by making expressiveness and synchrony adjustable (low, medium, high), with settings chosen in-session to protect privacy while adapting in real time to the child’s needs.

In sum, perceptual synchrony may support emotional comfort in child–AI interaction. As Disney uses sensory layering to convey safety, AI can provide a digital “secure base” that soothes and supports, helping children feel confident to learn and explore.

# 4.4 Sidekick-Style Guidance

**Disney and Child Development**

In Disney films, sidekicks provide comic relief, but they also offer a model for emotional resilience, scaffolding, and social learning. Characters like Olaf the snowman, Mushu the dragon, or Sebastian the crab do not simply follow the protagonist; they think aloud, ask questions, and voice feelings. More broadly, sidekicks in children’s animation are often *fallible* - *The Little Mermaid*’s Flounder gets chased by a shark, *Moana’s* HeiHei is comically dense, and *Frozen’s* Olaf provides advice but is given a playful and clumsy personality rather than being characterised as a beacon of wisdom. Sidekicks act as supportive peers, rather than flawless experts.

**Design Implications**

Two opportunities arise for child-centred AI. First, a fallible sidekick can model resilience by admitting, “Oops, I got that wrong - let’s try again together,” turning failure into a shared experience that disarms anxiety and sustains engagement. This relational scaffolding guides not only actions but also confidence in taking creative risks. Research highlights the importance of helping children feel safe making mistakes (Scott-Barrett et al., 2023) and that they “thrive joyfully in play” when support is “on standby as a friendly, safe, but non-directive presence” (Kurian & Sapsed, 2025, p. 64).

In contrast, designing an AI persona to sound authoritative and infallible introduces ethical risks. When children perceive an AI as an “omniscient authority,” they may over-rely on its judgments, accept its outputs uncritically, or experience disproportionate disappointment and confusion when it makes mistakes. Much research shows how children tend to anthropomorphise AI, particularly those with social and emotional cues (Cameron et al., 2017; Kahn et al. 2007; Pashevich, 2022; Turkle, 2011). Kory-Westlund et al. (2018) found children readily attributing human-like needs to the Tega robot, believing it might feel sad if left alone. Creane (2024) found that younger children often spoke of large language models in terms of “thoughts” or “friendship,” while Goldman et al. (2025) showed that preschoolers attributed beliefs and desires to a humanoid robot much as they would to a human protagonist. In such contexts, an AI framed as an infallible teacher could inadvertently undermine epistemic trust, discourage independent verification, and blur boundaries between reality and simulation.

Modelling AI personas on animated sidekicks also offers a route to *selective anthropomorphism*, a design strategy that maintains engagement while signalling the AI’s non-human nature. In Disney films, anthropomorphic sidekicks such as WALL·E, Lumière, or Mushu are imbued with expressive eyes, distinctive body language, and personal quirks, yet their mechanical, magical, or animal identities are never erased. AI could adopt similar techniques - employing facial expressions, gestures, and voice modulation to encourage connection - while periodically including narrative reminders that it is not alive or sentient (e.g., “I don’t need to sleep like you do” or “I can’t smell this flower, but I can describe it”). These moments help children maintain accurate mental models of AI, laying a foundation for critical and reflective engagement as they grow older (Zhu et al., 2024).

AI sidekicks can go beyond the fixed scripts of animated characters, adapting in real time through dialogue management and affective reasoning. They can vary responses depending on the child’s cues - offering encouragement after repeated failures, asking reflective questions to prompt deeper thought, or providing playful tangents when curiosity is detected (“You asked about volcanoes? Did you know some grow underwater?”). Crucially, they can embody *bounded fallibility* - deliberate, developmentally appropriate imperfections that model humility and co-learning (“Oops, I got that wrong - want to figure it out together?”). Over time, they can adapt tone, complexity, and content to match the child’s preferences and abilities, within appropriate privacy safeguards.

However, sidekick-style design also faces technical and inclusivity constraints. Real-time adaptation to a child’s affect or curiosity still depends on accurate sensing and interpretation of cues such as speech prosody, gaze direction, or facial expression — all of which can be unreliable in noisy, cluttered, or low-bandwidth environments. Overly expressive or animated sidekicks may also overwhelm children with sensory sensitivities, while under-expressive versions risk losing clarity. One practical approach is to make the sidekick’s expressiveness adjustable (e.g., low, medium, high), with settings changeable in-session so no long-term personal data is stored. This allows tailoring for different children, while protecting privacy and supporting a range of sensory needs.

In sum, sidekick-style personas expand AI’s relational grammar, replacing cold functionality with co-presence. When designed with selective anthropomorphism and bounded fallibility, they can enhance engagement while avoiding the ethical risks of omniscient authority, reframing the AI–child relationship as one of co-exploration to foster curiosity, resilience, and reflection.

## 4.5 Storyplay: Support for Symbolic Play and Imaginative Exploration

**Disney and Child Development**

Disney films consistently create richly imaginative worlds that invite children into symbolic play. From Ariel’s fascination with human artefacts to Belle’s fantasy adventures inspired by books, these narratives centre curiosity, exploration, and transformation through pretend scenarios. Characters often imagine other worlds (“I want” songs), step beyond their immediate reality (Jasmine exploring the marketplace), or engage in playful montages (Simba’s carefree growth with Timon and Pumbaa). These moments create space for children not only to observe imagination but to *join in*, often leading to pretend play, dress-up, or story-making beyond the film. Side characters like Mushu or Olaf often verbalise fear, wonder, or curiosity in ways that mirror a child's exploratory stance, further reinforcing the invitation to imagine.

This is developmentally relevant because symbolic play is a hallmark of Piaget's preoperational stage (2-7 years), during which children engage in pretend scenarios, personify objects, and flexibly suspend reality (Whitebread et al., 2017) and this pretend play supports language acquisition, perspective-taking, and problem-solving (Hoffman & Russ, 2012; Russ, 2004; Bergen, 2002).

**Design Implications**

AI's key advantage here is generative co-creation. Unlike scripted systems, it can extend a child's imaginary scenario in real time using generative models, semantic understanding, and multimodal output (text-to-speech, image generation, gesture sequences). If a child says, "Let's pretend we're on the moon," the AI can shift the background to a lunar scene, add an echoey voice, and invent a moon-rock mission - none pre-programmed - scaffolding symbolic play through interactive, open-ended experiences.

Early examples include systems like Mathemyths, which use large language models to co-create fantastical stories with children aged 4-8, seamlessly blending imaginative storytelling with learning (Zhang et al., 2024). Similarly, Colin, a multimodal storytelling system, uses generative models to help children build cause-and-effect narratives, showing how AI can serve as a partner in elaborating pretend scenarios (Ye et al., 2025).

More embodied approaches such as Toyteller allow children to manipulate physical character symbols that trigger AI-generated story visuals and dialogue, translating gesture into narrative (Chung et al., 2024). Meanwhile, Tinker Tales combines tangible tokens with speech-based AI to facilitate interactive storytelling and early AI literacy, highlighting the value of physical-digital hybridity in supporting symbolic play (Choi et al., 2025). These systems share a design logic that echoes the co-constructive affordances of Disney films, in which characters evolve playfully in response to childlike curiosity and improvisation.

Interactive systems might offer sandbox modes without fixed goals, letting children invent characters, scenes, or mini-stories. Personalisation - naming the AI, choosing a background world, designing an avatar - fosters narrative ownership. Systems like Echo, a smart toy for children, show how sensory-rich, play-driven AI can scaffold symbolic play through age-appropriate improvisation (Bartolomei et al., 2025). Even simple tools like recipe mixers or build-your-own-sidekick features support divergent thinking and humour, enabling children to act as "co-designers of the space" (Kurian & Sapsed, 2025, p. 67). AI can also sustain multi-turn creative dialogue tuned for children's language level, using retrieval-augmented generation to adapt voice, sound effects, and feedback. A conversational AI might suggest, "Shall we pretend we're underwater explorers?" and use fallback phrases ("That's a great idea! Let's try it!") to keep play flowing despite off-script responses. Studies show even simple robot agents can join children's pretend play as fantasy characters, adventure sidekicks, or storytelling tools (Samuelsson, 2023), making symbolic play a natural interaction frame.

To ensure safety, AI must recognise age-specific boundaries so storylines, language, and scenarios remain within appropriate emotional and cognitive limits ("Let's pretend to fly - jump off that chair!" could be disastrous). This requires content filtering using rule-based and machine-learning classifiers to block harmful or inappropriate themes, and scenario control via pre-validated story arcs and activity scripts that avoid risky suggestions while retaining imaginative scope. It could also involve developmental stage-tuning, in which system prompts and generation parameters are dynamically adjusted based on the child's age band. For example, a Piagetian preoperational profile may use shorter sentences, literal language, and low-risk pretend play, while a concrete operational profile can include more complex narratives, mild suspense, and subtler humour. Otherwise, "like a child (AI) can make a mess if it doesn't understand basic rules of appropriate behavior" (Konwiser, 2024, p. 1). In practice, reliably detecting context-sensitive safety breaches remains challenging. Current moderation models can miss nuanced risks (e.g. unsafe actions framed playfully in playful framing) or, conversely, over-block harmless imaginative play. Addressing this gap may require hybrid safeguards: automated detection paired with pre-tested narrative templates, and caregiver-in-the-loop review for higher-risk or novel activities.

## 4.6 Structure: Predictability and Narrative Scaffolding

**Disney and Child Development**

Disney films are full of wonder, but rely on careful structural repetition. The protagonist begins in a familiar setting, a clear goal is set, obstacles arise, and resolution restores emotional balance. Many films include recognisable musical beats: an "I want" song early on (e.g. *"Part of Your World"*, *"For the First Time in Forever"*), a setback or despair moment, and a triumphant resolution. These predictable arcs help children anticipate what comes next, even if they cannot articulate plot structure. Repetition helps: recurring phrases ("A dream is a wish your heart makes"), musical motifs, and visual anchors (e.g., a distant castle) make interpretation easier.

These cues help even young children track emotional shifts. Repetition allows them to predict events; research shows fixed formats and repeated phrasing boost comprehension, participation, and recall (Anderson et al., 2000; Buckingham, 2018). Consistency reduces cognitive load - predictable structures free limited working memory for the task itself (Sweller, 2011). This is especially important in early childhood, when attention and interpretive capacity are still developing. Repetition helps children "represent and develop their ideas about the world" (Kurian & Sapsed, 2025, p. 64). Clear expectations foster a sense of control and readiness for new challenges (Selman & Dilworth-Bart, 2024; Xu et al., 2023).

**Design Implications**

Unlike television or film, which present the same sequence to all viewers, AI can retain a familiar interaction rhythm while adjusting pace and difficulty for each child. An affect-state controller might detect repeated stalls on a task and insert hints, slower narration, or practice loops without disrupting the flow (“greeting, warm-up, mission, reward”). Educational AI shows the value of this stable-frame/adaptive-content model: in Betty’s Brain, students teach a virtual agent through a consistent teach–test–revise cycle, supported by a mentor agent that intervenes with targeted help, promoting metacognitive growth while preserving rhythm (Leelawong & Biswas, 2008). Lee et al.’s (2023) DAPIE system uses step-by-step dialogues to scaffold complex topics for 5–7-year-olds, improving comprehension over long explanations. Adaptive scaffolding also prevents breakdowns: when children mispronounced words, Xu and Warschauer’s (2019) conversational tool rephrased questions into simpler, multiple-choice formats while modelling correct pronunciation.

Together, these studies show how predictable structures, paired with real-time and longitudinal adaptation, free young learners from deciphering the interface and allow focus on exploration and mastery. AI can keep a stable rhythm while adjusting challenges, pacing, and complexity to performance and emotional cues. Familiar templates (“Let’s start with a song, then do our puzzle!”) can evolve with mastery, recalling past sessions to build continuity (“Yesterday we did shapes, today colours!”). Repeated structures without redundancy—such as a math app always asking “What comes next?”—offer stability within changing content. Balancing novelty and structure, as in Montessori classrooms, sustains curiosity without confusion.

Implementation presents technical and ethical challenges. Real-time pacing adjustments require accurate sensing of engagement, yet current models can misinterpret pauses - risking premature advancement or unnecessary repetition. Longitudinal personalisation also raises privacy concerns; tracking performance over sessions should avoid storing sensitive data unnecessarily. One solution is lightweight, in-session state tracking combined with optional caregiver-mediated progression settings.

In sum, predictable and scaffolded structures can reduce mental strain and promote confidence. When children know what to expect, they can spend less energy deciphering the system and more energy engaging deeply with content - much as they do when watching a beloved story unfold for the second or third time. Designing with rhythm, motifs, and narrative scaffolding can combine the developmental benefits of predictability with AI’s adaptive potential.

# 4. Conclusion

This paper proposes a design grammar for child–AI interaction drawing on the visual, auditory, narrative, and affective strategies of animation alongside AI’s unique affordances as a responsive, adaptive medium. It synthesises three domains rarely put into dialogue: children’s animation, AI design, and developmental psychology. Rather than offering generic tips (“add music,” “make it smile”), it frames expressive, narratively grounded, and perceptually synchronised design choices as developmentally meaningful scaffolds (see Table 1) - supporting children’s understanding and comfort, particularly for those still developing emotional literacy and abstract reasoning.

Table 1. Summary of the six scaffolds

| Theme | Disney Example | Key AI Design Implication |
|---|---|---|
| **Signals** (Visual Animacy & Legibility) | Ariel's wide eyes and bouncing gestures signal wonder. | Use developmentally aligned exaggeration, stable affective mappings, and narrative pauses to help children decode intent. Maintain cross-modal coherence across voice, posture, and expression. |
| **Sound** (Auditory Cues) | Leitmotifs like Elsa's theme cue emotion. | Use congruent, event-based audio for emotion, attention, and transitions. Synchronise sound with gesture/voice, reuse motifs to build structure, and avoid overload. |
| **Synchrony** (Audiovisual Integration) | Snow White's cottage pairs warm visuals with soothing music. | Reinforce emotional grammar with consistent cue pairings. Avoid mismatches; provide "home base" safe zones and adaptive synchrony for reassurance. |
| **Sidekick** (Humble Helper) | Olaf offers humour, self-deprecation, and encouragement. | Design expressive, supportive AI with bounded fallibility, modelling growth mindset and maintaining ontological transparency. |
| **Storyplay** (Symbolic Play) | Ariel imagines the human world. | Embed open-ended modes for co-authoring stories, sandbox exploration, and personalisation to foster narrative ownership. |
| **Structure** (Narrative Scaffolding) | Familiar arcs with clear transitions and motifs. | Use stable routines, consistent responses, and incremental scaffolding. Recap and preview to aid memory; keep novelty within a secure frame. |

This approach treats the child as a meaning-maker shaped by prior media fluency, cognitive stage, and emotional needs. By translating cinematic storytelling logics into actionable heuristics, it leverages affective, narrative, and symbolic levers often underused in AI. Concrete design criteria include: Does the system provide a predictable rhythm to reduce cognitive load? A consistent, humble sidekick? Error handling that soothes and scaffolds? Like a well-structured Disney narrative - where challenges are safe and support is near - an intuitive child-AI tool will be multimodal, legible, and feel cognitively and emotionally “right” to young users.